\newcommand{\ifJCAP}[2]{#2} 
\ifJCAP{
  \documentclass[10pt]{iopart}
  \newcommand{\text}[1]{{{\rm #1}}}
  \usepackage[colorlinks,
              linkcolor=blue,
              citecolor=blue,
              ]{hyperref}
}
{ 
  \documentclass[aps,prd,10pt,
                 preprintnumbers,numbers,sort&compress,nofootinbib,
                 twocolumn,
                 showpacs,
                 colorlinks,
                 linkcolor=blue,
                 citecolor=blue]{revtex4}
  \usepackage{amsmath}
}
\usepackage{amssymb}

\newcommand{\vect}[1]{\vec{\mathbf{#1}}}
\newcommand{\QCD}{\textsc{qcd}}

\ifJCAP{
  \makeatletter
  \def\@fnsymbol#1{^{\thefootnote}\relax}
  \setcounter{footnote}{0}
  \makeatother
}
{ 
}
\newcommand{\exclude}[1]{}

\newcommand{\Chandra}{\textit{Chandra}}
\newcommand{\SPI}{\textsc{spi}}
\newcommand{\CGRO}{\textsc{cgro}}
\newcommand{\INTEGRAL}{\textsc{integral}}
\newcommand{\EGRET}{\textsc{egret}}
\newcommand{\DM}{\textsc{dm}}
\newcommand{\keV}{\text{ keV}}
\newcommand{\GeV}{\text{ GeV}}
\newcommand{\MeV}{\text{ MeV}}

\newcounter{myenumi}
\newenvironment{myenumerate}[1][]{
  \begin{list}{(#1\arabic{myenumi})}{
      \usecounter{myenumi}
      \setlength\parsep{0pt}
      \setlength\itemsep{0pt}
      \setlength\itemindent{0pt}
      \setlength\topsep{0pt}
      \setlength\listparindent\parindent
      \setlength\labelsep\leftmargin
      \addtolength\labelsep{-\labelwidth}
      \addtolength\labelsep{\itemindent}
    }
}{\end{list}}

\ifJCAP{
  \newcommand{\keywords}[1]{\noindent \textit{Keywords}\/ #1}
  \newcommand{\preprint}[1]{}
  \newcommand{\acknowledgments}{\ack}
}
{ 
}
\begin{document}
\title{Diffuse x-rays: Directly Observing Dark Matter?}
\ifJCAP{
  \author{Michael McNeil Forbes$^{1}$ and Ariel R. Zhitnitsky}
  \address{$^1$ Nuclear Theory Group, Department of Physics, University
    of Washington, Seattle, WA 98195-1560, USA} 
  \ead{\mailto{mforbes@alum.mit.edu}}  
  \address{$^2$ The University of British Columbia, Department of Physics
    and Astronomy, Vancouver, B.C., V6T 1Z1, Canada}
}
{ 
  \author{Michael McNeil Forbes}
  \affiliation{Nuclear Theory Group, Department of Physics, University
    of Washington, Seattle, WA 98195-1560, USA} 
  \author{Ariel R. Zhitnitsky}
  \affiliation{The University of British Columbia, Department of Physics
    and Astronomy, Vancouver, B.C., V6T 1Z1, Canada}
}
\begin{abstract}
  Several independent observations of the Galactic core suggest
  hitherto unexplained sources of energy.  We suggest that dark matter in
  the form of dense antimatter nuggets could provide a natural site
  for electron and proton annihilation, providing 511\keV\ photons,
  gamma-rays, and diffuse\keV\ x-ray radiation.  We show that
  identifying dark matter as antimatter nuggets is consistent with the
  observed emissions, and we make definite predictions about their
  spectrum and morphology.  If correct, our proposal not only
  identifies dark matter and explains baryogenesis, but allows x-ray
  observations to directly probe the matter distribution in our
  Galaxy.
\end{abstract}
\preprint{NT@UW-06-25}
\maketitle
\section{Introduction:}
Two of the outstanding cosmological mysteries---the natures of dark
matter and baryogenesis---might be explained by the idea that dark
matter consists of Compact Composite Objects
(\textsc{cco}s)~\cite{Zhitnitsky:2002qa,Oaknin:2003uv,Zhitnitsky:2006vt},
similar to Witten's strangelets~\cite{Witten:1984rs}.  The idea is
that these \textsc{cco}s---nuggets of dense matter and
antimatter---form at the same \textsc{qcd} phase transition as
conventional baryons (neutrons and protons), providing a natural
explanation for the similar scales $\Omega_{\DM} \approx
5\Omega_{\textsc{b}}$.  Baryogenesis proceeds through a charge
separation mechanism: both matter and antimatter nuggets form, but the
natural \textsc{cp} violation of the so-called $\theta$ term in
\textsc{qcd}\footnote{If $\theta$ is non-zero, one must confront the
  so-called strong \textsc{cp} problem whereby some mechanism must be
  found to make the effective $\theta$ parameter extremely small today
  in accordance with measurements.  One of the most natural
  resolutions is through a dynamical axion.  Domain walls associated
  with this field (or ultimately, whatever mechanism resolves the
  strong \textsc{cp} problem) play an important role in forming these
  nuggets, and may play in important role in their ultimate stability.
  See~\cite{Zhitnitsky:2002qa,Oaknin:2003uv} for details.} drives the
formation of more antimatter nuggets than matter nuggets, resulting in
the left-over baryonic matter that forms visible matter today
(see~\cite{Oaknin:2003uv} for details).  The idea that \textsc{cp}
violation may be able to effectively drive charge separation may
already have found experimental support through the Relativistic Heavy
Ion Collider (\textsc{rhic}) at Bookhaven \cite{Kharzeev:2007tn}.

This mechanism requires no fundamental baryon asymmetry to explain the
observed matter/antimatter asymmetry: $B_{\text{univ}}=B_{\DM} +
B_{\text{Visible}}-\bar{B}_{\DM}=0$ where $B_{\DM}$($\bar{B}_{\DM}$)
is the total (anti)baryon charge contained in the dark (anti)matter
nuggets.  From this and the observed relation $\Omega_{\DM} \approx 5
\Omega_{\textsc{b}}$ we have the approximate ratios
$\bar{B}_{\DM}$:$B_{\DM}$:$B_{\text{Visible}}\simeq $ 3:2:1.

Unlike conventional dark matter candidates, dark antimatter nuggets
would be strongly interacting, but macroscopically large objects, and
would not contradict any of the many known observational constraints
on dark matter or
antimatter~\cite{Zhitnitsky:2006vt,Forbes/Zhitnitsky:2007b} for three
reasons:
\begin{myenumerate}
\item They carry a huge (anti)baryon charge $|B| \approx
  10^{20} \text{--} 10^{33}$, so they have an extremely tiny number
  density.  This would explain why they have not been directly
  observed on earth.  The local number density of dark matter
  particles with these masses is small enough that interactions with
  detectors are exceedingly rare and fall within all known detector
  and seismic constraints~\cite{Zhitnitsky:2006vt}.
\item They have nuclear densities, so their interaction cross-section
  is small $\sigma/M \approx 10^{-13}\text{--}10^{-9}$~cm$^2$/g.  This is
  well below typical astrophysical and cosmological limits which are
  on the order of $\sigma/M<1$~cm$^2$/g.  Dark-matter--dark-matter
  interactions between these nuggets are thus negligible.
\item They have a large binding energy such that baryons in the
  nuggets are not available to participate in big bang nucleosynthesis
  (\textsc{bbn}) at $T \approx 1$\MeV.  In particular, we suspect that
  the baryons in these nuggets form a superfluid with a gap of the
  order $\Delta \approx 100$\MeV.  This scale would provide a natural
  explanation for the observed photon to baryon $n_{B}/n_{\gamma} \sim
  10^{-10}$, which requires that formation of the nuggets stop at
  precisely $T_{\text{formation}} =
  41$\MeV~\cite{kolb94:_early_univer,Oaknin:2003uv}.  At temperatures
  below this, incident baryons with energies below the gap would
  Andreev reflect rather than become incorporated into the
  nugget.\footnote{If this mechanism is verified, then this value of
    $T_{\text{formation}}$ would provide one of the most accurate
    constraints on the properties of high-density \QCD\ matter.}
\end{myenumerate}
Thus, on large scales, the nuggets would be sufficiently dilute that
they would behave as standard collisionless cold dark matter
(\textsc{ccdm}).  When the number densities of both dark and visible
matter become sufficiently high, however,
dark-antimatter--visible-matter collisions may release significant
radiation and energy.

In this paper, we explain how the interaction between visible matter
and dark antimatter nuggets in our Galaxy could naturally resolve
several outstanding observational puzzles.  These annihilations in the
galactic core may be the primary source of detected 511\keV\
radiation, and the primary source of energy sustaining diffuse\keV\
emissions.  We consider three independent observations of diffuse
radiations from the galactic core:
\begin{myenumerate}
\item \SPI/\INTEGRAL\ observes 511\keV\ photons from positronium decay
  that is difficult to explain with conventional astrophysical
  positron
  sources~\cite{Knodlseder:2003sv,Beacom:2005qv,Yuksel:2006fj}.  Dark
  antimatter nuggets would provide an unlimited source of positrons as
  suggested in~\cite{Oaknin:2004mn,Zhitnitsky:2006tu}.
\item \Chandra\ observes a diffuse\keV\ x-ray emission that greatly
  exceeds the energy from identified sources~\cite{Muno:2004bs}.
  Visible-matter/dark-antimatter annihilation would provide this
  energy.
\item \EGRET/\CGRO\ detects\MeV\ to\GeV\ gamma-rays, constraining
  antimatter annihilation rates.
\end{myenumerate}
The main goal of this paper is to demonstrate that these observations
suggest a common origin for both the 511\keV\ radiation and the
diffuse\keV\ x-ray emission.  Assuming dark antimatter nuggets to be
the common source, we extract some phenomenological parameters
describing their properties.  We show that dark antimatter is
consistent with all of these data, and that it may fully explain the
missing sources of emission without requiring the invention of new
non-baryonic fields for dark matter.  Finally, we argue that the
spectrum of the\keV\ emission should be independent of position, and
that there should be a spatial correlation between the various
emissions and the distributions of dark and visible matter.  These
concrete predictions should allow our proposal to be verified or ruled
out by observations and analysis in the near future.

Our purpose here is not to discuss details of a particular model for
the nuggets, but to show generally that macroscopic antimatter nuggets
could easily explain the missing energy and missing positrons in the
core of our Galaxy.  To this end, we postulate a few basic properties
to be discussed in detail elsewhere~\cite{Forbes/Zhitnitsky:2007b}:
\begin{myenumerate}
\item[(A.1)] The antimatter nuggets provide a virtually unlimited source of
  positrons (e$^{+}$) such that impinging electrons (e$^{-}$) will
  readily annihilate at their surface through the formation of
  positronium~\cite{Oaknin:2004mn,Zhitnitsky:2006tu}.
\end{myenumerate}
About a quarter of the positronium annihilations release back-to-back
511\keV\ photons.  If this reaction occurs on the surface, on average
one of these photons will be absorbed by the nugget while the other
will be released.
\begin{myenumerate}
\item[(A.2)] The nuggets provide a significant source of anti-baryonic matter
  such that impinging protons will annihilate.  We assume that the
  proton annihilation rate is directly related to that of
  electrons through a suppression factor $f<1$ to be discussed below.
\end{myenumerate}
Proton annihilation events will release about $2m_{p} \approx 2$\GeV\
of energy per event and will occur close to the surface of the nugget
creating a hot spot that will mainly radiate x-ray photons with\keV\
energies rather than\GeV\ gamma-rays, (see the appendix for a detailed
discussion.)  To connect these emissions with observations, we make an
additional assumption:
\begin{myenumerate}
\item[(A.3)] We assume that these emitted 511\keV\ photons dominate the
  observed 511\keV\ flux and that these proton annihilations provide the
  dominant source of energy fueling the\keV\ emission.
\end{myenumerate}
This allows us to use the 511\keV\ observations to estimate the
electron annihilation rate and the \Chandra\ flux to estimate the
proton annihilation rate.  This assumption may not be fully
realized---the observed fluxes may be a combination of various
conventional and unconventional sources---but by assuming that the
majority of the flux originates from matter-antimatter annihilations,
we are able to provide constraints on the annihilation properties of
the nuggets.  As we shall discuss, these constraints, although
only bounds, are non-trivial and could have ruled out our proposal.

\section{Proposal:}
The primary feature of our proposal is that the antimatter nuggets
provide a single annihilation target for both electrons and protons.
As a result, both the 511\keV\ emission from electron annihilation and
the x-ray emission from proton annihilation should originate from the
same regions of space, with the local rates of annihilation
proportional to the product of the number densities of visible and
dark matter $n_{V}(\vect{r})n_{D}(\vect{r})$.

By comparing several different observational sources, one may remove
the dependence on the dark and visible matter distributions because
all observations integrate the radiation along the same line of sight
from the earth to the core of the Galaxy.  We make the approximation
that the relevant line-of-sight average is the same for the three sets
of data we consider.  As a result, direct comparisons between the data
provide non-trivial insights about the properties of dark antimatter,
independent of the matter distributions.
 
\subparagraph{\Chandra:} X-ray emissions from the galactic core
provide a puzzling picture: they seem to indicate that an 8\keV\
thermal plasma is being maintained, but the source of energy fueling
this plasma is a mystery.  After subtracting known x-ray sources from
the \Chandra\ x-ray images of the galactic core, one finds a residual
diffuse thermal x-ray emission with a thermal component well described
by a hot 8\keV\ plasma with surface brightness $\Phi_{\Chandra} =
(1.8\text{--}3.1)\times 10^{-6}$~erg/cm$^2$/s/sr~\cite{Muno:2004bs}.  To
sustain such a plasma would require some $10^{40}$~erg/s of energy in
the galactic core, which is much more than the observed rate of
supernovae, for example, can explain~\cite{Muno:2004bs}.

Protons annihilating with dark antimatter would release some $2m_{p}
\approx 2$\GeV\ of energy per event.  Assuming (A.3.) this to be the
dominant source of energy sustaining the\keV\ emission, one may relate
the x-ray luminosity to the number of annihilations events observed
along the line of sight.  Proton/dark-antimatter annihilations in the
galactic core must thus provide an energy input of
\begin{equation}
  \label{eq:phiT}
  \Phi_{\Chandra} \approx 2\times 10^{-6}
  \frac{\text{erg}}{\text{cm}^2\cdot \text{ s}\cdot \text{ sr}}
\end{equation}
into\keV\ photons along our line of sight as observed
in~\cite{Muno:2004bs}.  We shall see that our proposal can easily
accommodate this flux.

\subparagraph{SPI/INTEGRAL:} An observed flux of 511\keV\ photons
from the core of the Galaxy is another puzzle.  It is clear that this
comes from low-energy e$^{-}$e$^{+}\rightarrow2\gamma$ annihilation
events (positronium decay), but the source of positrons is a mystery
(the interstellar medium contains many electrons).  After accounting
for known positron sources, only a small fraction of the emission may
be explained~\cite{Knodlseder:2003sv,Beacom:2005qv,Yuksel:2006fj}.
\SPI\ has measured the flux of 511\keV\ photons from the galactic
bulge to be $9.9^{+4.7}_{-2.1}\times 10^{-4}$~ph/cm$^2$/s with a half
maximum at
$9^\circ$~\cite{Knodlseder:2003sv,Jean:2003ci,Knodlseder:2005yq,Jean:2005af}.

As discussed in~\cite{Oaknin:2004mn,Zhitnitsky:2006tu}, dark
antimatter nuggets provide a natural source of positrons in the
``atmosphere'' of positrons (electrosphere) surrounding the nuggets
that will readily annihilate with electrons from the \textsc{ism},
(see equation (\ref{eq:field}) in the appendix).  The resulting spectrum
will be dominated by the formation and subsequent decay of
positronium, one quarter of which will decay from the $^{1}$S$_{0}$
state giving rise to two back-to-back 511\keV\ photons.  Under
assumption A.1., on average, one of these photons will be radiated.
These $^{1}$S$_{0}$ annihilations account for approximately one
quarter of the annihilations events: the rest occur through the
$^{3}$S$_{1}$ state and give rise to a continuum below 511\keV\ that
has also been measured~\cite{Weidenspointner:2006nu}, confirming the
source as positronium decay.

Assuming (A.3.)\ that this e$^{-}$e$^{+}$ annihilation is the dominant
source of the measured 511\keV\ photons, the flux from the core of the
Galaxy provides an estimate of the rate of annihilation events.
Unfortunately, the angular resolution of \INTEGRAL\ is not very precise.
To estimate the core flux requires some assumption about the spatial
distribution of the annihilations.  We simply use the estimates
in~\cite{Boehm:2003bt,Hooper:2003sh} to obtain an order of magnitude
for same region observed by \Chandra\  (\ref{eq:phiT}).
Multiplying by a factor of 4 to account for the continuum emission,
the rate of annihilations integrated along our line of sight is
\begin{equation}
  \label{eq:phiee}
  \phi_{\text{e}^{-}\text{e}^{+}} \approx 
  10^{-1} \frac{\text{event}}{\text{cm}^2\cdot \text{ s}\cdot \text{ sr}}.
\end{equation}
\subparagraph{EGRET:} Some fraction of the proton annihilation events
will produce high energy photons that may directly escape.  These
photons, with energies up to 1\GeV, should be detected by \EGRET\ aboard
the \CGRO\ satellite, which has measured the intensity of photons from
30\MeV\ to 4\GeV\ from a central core region of $60^\circ$ longitude and
$10^\circ$ latitude~\cite{Strong:2004de,Kamae:2004xx}.  Much of this
emission is due to cosmic rays interacting with visible matter.  We
therefore define the parameter $a_{\EGRET} < 1$ to be the
fraction of the total \EGRET\ intensity due to direct photon production
from proton/dark-antimatter annihilations.

We may thus use the \EGRET\ data to provide two constraints on
proton/dark antimatter annihilations.  Performing the energy integrals
from 100\MeV\ to 1\GeV\ (the typical range of photon energy
from $\text{p}^{+}\text{p}^{-}$ annihilations) on the data
from~\cite{Kamae:2004xx}, we obtain the following integrated count
$\phi_{\text{p}^{+}\text{p}^{-}\rightarrow\gamma}$ and energy
$\Phi_{\EGRET}$ surface brightnesses extracted from the \EGRET\ data:
\ifJCAP{
  \begin{eqnarray}
    \label{eq:egret}
    \phi_{\text{p}^{+}\text{p}^{-}\rightarrow \gamma} &\approx& 10^{-4}
    \frac{\text{ph}}{\text{cm}^2\cdot  \text{ s} \cdot \text{ sr}},\label{eq:phipp}\\
    \Phi_{\EGRET} &\approx& 10^{-7}
    \frac{\text{erg}}{\text{cm}^2\cdot \text{ s} \cdot \text{ sr}}.\label{eq:phiE}
  \end{eqnarray} 
}{ 
  \begin{subequations}
    \label{eq:egret}
    \begin{align}
      \phi_{\text{p}^{+}\text{p}^{-}\rightarrow \gamma} &\approx  10^{-4}
      \frac{\text{ph}}{\text{cm}^2 \text{ s}\text{ sr}},\label{eq:phipp}\\
      \Phi_{\EGRET} &\approx 10^{-7}
      \frac{\text{erg}}{\text{cm}^2\cdot \text{ s} \cdot\text{ sr}}.
      \label{eq:phiE}
    \end{align}
  \end{subequations}
}
\section{Comparisons:}
We now can compare (\ref{eq:phiT}), (\ref{eq:phiee}), and
(\ref{eq:egret}) to extract some properties of the visible
matter/dark-antimatter annihilation rates.
\subsection{\Chandra\ Constraints:}
Proton-antiproton annihilation should proceed at a rate
$\phi_{\text{p}^{+}\text{p}^{-}} = f\phi_{\text{e}^{-}\text{e}^{+}}$
proportional to, but different from, that for electron-positron
annihilation.  The suppression factor $f$ is due to several effects.
\begin{myenumerate}
\item The most important is that the transition from quark-matter to
  vacuum occurs on a \QCD\ scale and is very sharp: $\sim\text{fm}\sim
  (200\;\MeV)^{-1}$.  This needs to be compared with the long
  de-Broglie wavelength of the incoming protons, which have momenta on
  the scale $p=m_{p}v \sim $\MeV.  As such, the effective potential
  for the incoming protons due to this transition will behave
  qualitatively as a delta-function and the reflection probability
  will be high, irrespective of whether the interaction is attractive
  or repulsive.  (This is a standard and general result from
  one-dimensional quantum mechanics.)
\item Additional suppression may result from superfluid correlations
  in the nugget.  This would lead to the analogue of Andreev
  reflection in superconductors where incident protons reflect, rather
  than penetrate.
\item Finally, charge exchange reactions may take places which allow
  the nuggets to maintain overall neutrality charge neutrality,
  despite the higher rate of positron annihilation than proton
  annihilation.
\end{myenumerate}
Together these effects lead to a suppression factor $f$ which, in
principle, may be calculated.  The result, however, will be highly
model dependent.  Instead, we introduce this as a phenomenological
parameter.  Note that this parameter must be less than unity: $f<1$.
If observations determine this parameter to be large, the model as a
whole can be ruled out.  Thus, even without a precise calculation of
$f$, observations provide a nontrivial test of our proposal.  Once
$f$ is known, one can estimate the rate at which energy is deposited
into the core of the Galaxy.

Our picture is that the fraction $f$ of protons that annihilate
penetrate a short way into the dark antimatter nuggets before
annihilation.  Annihilation produces a handful of high energy
particles, (mostly Goldstone bosons).  About half of these penetrate
deep withing the nuggets, ultimately thermalizing and radiating
low-energy photons (this emission is not subject of the present work)
but the remaining half stream toward the surface of the nugget,
ultimately cascading into hundreds of highly energetic positrons that
stream into the electrosphere.  These positrons are accelerated by the
strong electromagnetic fields present at the surface of the nuggets
and radiate some fraction $g$ of the total 2\GeV\ annihilation energy
through Bremsstrahlung radiation with a typical\keV\ energy scale set
by the plasma frequency of the photons in the electrosphere.\footnote{
  The factor $1-g$ is introduced to account for the fraction of the
  2\GeV\ annihilation energy not detected by \Chandra.  This includes
  the half of the energy lost geometrically deep into the nugget
  producing very low energy photons, as well as the radiated energy in
  the form of very hard x-rays ($> 10\keV$) not detected by \Chandra.
  Such hard x-rays may in principle be emitted by this mechanism as
  discussed in the appendix.  It also accounts for other energy
  losses, such as neutrino emissions, but these do not play a dominant
  role.}

A detailed account of this process is presented in the appendix.
Note: we expect that this mechanism of radiation---where energetic
electrons are accelerated by a strong electric field present in the
electrosphere---is also the dominant mechanism for radiation in
strange quark stars, yet it appears to have been overlooked in the
literature of radiation from the surface of strange quark
stars~\cite{Prakash-Page-Usov-PC:2007}.

The end result is that each proton annihilation produces a ``hot
spot'' on the surface of the nugget that emits a broad spectrum of
photons on the 10\keV\ scale containing a total energy of $g\times
2$\GeV.  Although the details of this spectrum are difficult to
calculate and are model dependent, (see the discussion in the
appendix), one immediately has a strong prediction that the spectrum
should be relatively independent of the location of the emission.
This is in marked contrast with thermal plasma emission where the
spectrum depends sharply on the temperature of the plasma, which in
turn depends strongly on local parameters, such as the rate of energy
input and particle density in the surrounding
environment.\footnote{There will be a mild dependence of the spectrum
  on the thermal distribution of the incident protons, but this is a
  small effect as typical interstellar temperatures are much below the
  $10$\keV\ scale of these emissions.}  We note that this is already
consistent with analyses of the excess \EGRET\
emission~\cite{deBoer:2004ab}.

Our proposal would thus resolve the problem of heating an 8\keV\
thermal plasma in the core of the Galaxy by replacing the plasma with
an emitting hot-spot on the antimatter nugget surface fueled by proton
annihilation.  The similarity to a thermal spectrum arises from
Bremsstrahlung emission in the electrosphere of the nuggets stimulated
by the strong electric fields present near the surface of quark matter
objects
systems~\cite{Alcock:1986hz,Kettner:1994zs,Madsen:2001fu,Usov:2004iz}.
This is discussed further in section~\ref{sec:thermal-bremsstralung}
of the appendix.

The x-rays detected by \Chandra\ allow us to directly compare the
e$^{-}$e$^{+}$ annihilation rate (\ref{eq:phiee}) with the
observed\keV\ emission $\Phi_{\Chandra} = (2\text{
  \GeV})\,f\,g\,\phi_{\text{e}^{-}\text{e}^{+}}$.  From this we may
estimate the total suppression fraction $f\,g$ from (\ref{eq:phiT})
and (\ref{eq:phiee}):
\begin{equation}
  \label{eq:f}
  f\,g = \frac{\Phi_{\Chandra}}{(2\text{\GeV})\phi_{\text{e}^{-}\text{e}^{+}}}
  \approx 6\times 10^{-3}.
\end{equation}
As discussed in the appendix, we expect the suppression factor $g$ to
on the order of $\frac{1}{10} < g < \frac{1}{2}$.  The remaining
suppression factor $f$ could easily be accounted for by strong
interface effects at the surface of dark antimatter nuggets as
mentioned above.  The parameter $f$, however, is not entirely free: It
would be difficult, for example, to explain a value of $f>1$ greater
than unity due to the suppression factors discussed above.  Thus,
(\ref{eq:f}) provides a nontrivial test of our proposal that could
rule it out if strong arguments could be given that $f$ must be
greater than unity.  In principle, $f$ and $g$ are calculable from
detailed models of the nuggets, providing a direct test of any
quantitative model for dark antimatter.
\subsection{EGRET Constraints}\label{sec:egret-constraints}
We now consider the \EGRET\ constraints.  If proton annihilations
occur inside of the antimatter nuggets, then most of the decay
products will be strongly interacting and will scatter into positrons
or thermalize as discussed above. Occasionally, however, the
annihilation will directly produce a high-energy photon with an energy
of 100\MeV\ to 1\GeV\ that can easily escape the nugget.  These
processes must be compared with similar processes involving gluons rather
than photons.  Direct photon production is thus suppressed by a factor
of $\alpha/\alpha_{S}\sim 1/40\text{--}1/100$ due to the relative
weakness of the electromagnetic interaction compared to the strong
interaction.  Thus, we may convert the rate of proton annihilations to
a rate of direct photon emission along our line of sight, and compare
the integrated \EGRET\ data to the proton-antiproton annihilation
rates $a_{\EGRET}\,\phi_{\text{p}^{+}\text{p}^{-}\rightarrow \gamma} =
(\alpha/\alpha_{S})\,\Phi_{\Chandra}/(2\text{\GeV}) =
f\,(\alpha/\alpha_{S})\,\phi_{\text{e}^{-}\text{e}^{+}}$.
Using~(\ref{eq:phiT}) and~(\ref{eq:phipp}) we obtain
\begin{equation}
  \label{eq:aEGRET_comp_N}
  a_{\EGRET} = \frac{\alpha}{\alpha_{S}}\,
  \frac{\Phi_{\Chandra}/(2\text{\GeV})}
  {\phi_{\text{p}^{+}\text{p}^{-}\rightarrow \gamma}}
  \approx 10^{-1},
\end{equation}
where $a_{\EGRET}$ is the fraction of the total \EGRET\ intensity
resulting from this direct photon emission.  (Note: this estimate does
not include photons from neutral meson decays such as $\pi \rightarrow
2\gamma$ because these photons cannot escape the nuggets.  This is
discussed in detail in~\ref{sec:energy-transfer}.)  Finally, we may
compare the amount of energy released as high energy photons with the
integrated \EGRET\ energy flux $a_{\EGRET}\,\Phi_{\EGRET} =
(\alpha/\alpha_{S})\,\Phi_{\Chandra}$ to obtain an independent
estimate of $a_{\EGRET}$ from~(\ref{eq:phiT}) and~(\ref{eq:phiE})
\begin{equation}
  \label{eq:aEGRET_comp_E}
  a_{\EGRET} =
  \frac{\alpha}{\alpha_{S}}\frac{\Phi_{\Chandra}}{\Phi_{\EGRET}}
  \approx 3\times 10^{-1}.
\end{equation}
This is consistent with~(\ref{eq:aEGRET_comp_N}) to within the errors
of our approximations, and demonstrates the importance of the\GeV\
energy scale for both the x-rays band measure by \Chandra\ and the
gamma ray band measured by \EGRET.  This ``accidental'' coincidence
is naturally provided by the \QCD\ scale in our proposal.  Together,
equations~(\ref{eq:aEGRET_comp_N}) and~(\ref{eq:aEGRET_comp_E})
suggest that visible-matter/dark-antimatter annihilations could
account for a small fraction ($\sim 10\%$) of the measured \EGRET\
intensity.  Much of the \EGRET\ intensity has been accounted for, and
future improvements in this accounting could provide a tight
constraint on our proposal.  It is nontrivial that the measured
fraction $a_{\EGRET} <1$ be less than one: If the observations showed
$a_{\EGRET}$ to be greater than unity, then one would have had a
strong case against our proposal.\footnote{There is still some debate
  about the fraction of the observed \EGRET\ emission that can be
  ``explained'' with conventional backgrounds.  For example,
  \cite{deBoer:2004ab} suggests that the observations support an
  excess that may result from \textsc{wimp} dark-matter
  annihilations, while \cite{Strong:2004de} suggests that much of this
  could be explained by cosmic rays.  Our point is that the present
  constraints still leave a large region to accommodate our proposal.
  In principle, however, as the backgrounds are better understood,
  they will place stronger and stronger constraints on $a_{\EGRET}$,
  possibly ruling out our proposal.}

We have made several assumptions in our estimates that must be cleaned
up in order to make quantitative predictions.  We have assumed that
the line-of-sight integral over the matter distribution is the same
for each observation, and that the phenomenological suppression
factors $f$, $g$, as well as the fraction $a_{\EGRET}$, are independent
of the local environment and hence factor from this integral.  In
reality these factors will depend on the energies of the events, the
nature of the environment (ionized vs. neutral), the local velocities
of the visible and dark matter, and a host of similar factors.  We may
lump these factor together for any given line-of-sight, but it will be
important to quantify how these factors vary as different
line-of-sight averages are performed in order to make quantitative
comparisons along different lines of sight.

We have also purposefully neglected any particular models of the
matter distribution in our Galaxy.  By comparing several observations
along similar lines of sight, the results presented in this work do
not depend on details such as the typical baryon charge $B$ of the
nuggets, nor do they depend on the exact matter distributions.
Postulating a distribution leads to a direct connection between the
observed fluxes and properties of the dark antimatter nuggets such as
the interaction cross-section $\sigma/M \approx
10^{-13}\text{--}10^{-9}$~cm$^2$/g.  This is consistent with all known
observations and interaction constraints, but a more careful analysis
could significantly narrow the window of parameters.  We shall discuss
these and other details about the nature of dark antimatter
elsewhere~\cite{Forbes/Zhitnitsky:2007b}.

\section{Predictions:}
We have shown that the observed excess 511\keV\ photons and thermal
x-ray radiation from the core of our our Galaxy may be directly
explained by the hypothesis that much of the dark matter in the core
of our Galaxy is in the form of dark antimatter nuggets.  In this
case, both observations have the same physical origin and are
naturally correlated.  The observations suggest very reasonable values
for $f$ and $g$---the suppression of the proton to electron
annihilation rates and fraction of energy emitted---and $a_{\EGRET}$---the
fraction of \EGRET\ photons arising from dark matter annihilations.
These parameters satisfy the nontrivial constraints $f<1$ and
$a_{\EGRET}<1$, either of which could rule out our proposal.

Our proposal also makes the definite prediction that the 511\keV\ flux
and the thermal x-ray flux should be correlated.  For example,
\Chandra\ has detected a diffuse x-ray emission with flux $6.5\times
10^{-11}$~erg/cm$^{2}$/s/deg$^{2}$ from a region of the disk $28^\circ$
off the center~\cite{Ebisawa:2005zt}.  This is one order of magnitude
smaller than~(\ref{eq:phiT}), and so the integrated positronium
annihilation flux should be similarly reduced from~(\ref{eq:phiee}):
$\phi_{\text{e}^{-}\text{e}^{+}} = 10^{-2}$ event/cm$^{^2}$/s/sr.  We
predict that the local 511\keV\ flux should be one quarter of this
rate.

These fluxes should also be correlated with the product of the
distributions of dark and visible matter
$n_{D}(\vect{r})n_{V}(\vect{r})$.  The angular resolution of \INTEGRAL\
is not spectacular, but sufficient that the different source
distributions can be compared.  For example, as discussed
in~\cite{Frere:2006hp}, the measured angular distribution is
consistent spatial distribution models of
$n_{V}(\vect{r})n_{D}(\vect{r})$.  Unlike the mechanism suggested
in~\cite{Frere:2006hp}, however, our proposal can also easily account
for the observed luminosity.

Checking the angular 511\keV\ distribution with the matter
distribution inferred from the \Chandra\ x-ray data should provide a
nontrivial confirmation or refutation of this model.  One also has
the possibility of looking for correlations elsewhere from clusters or
in places where dark matter halo clumps pass through the disk.  The
problem here becomes one of sensitivity: \INTEGRAL\ is not sensitive
enough to detect the 511\keV\ emissions from extra-galactic sources.
\Chandra, however, may still be used to detect extra-galactic x-ray
emission.  If our proposal is correct, diffuse x-ray measurements with
subtracted sources could be directly translated into measurements of
the integrated $n_{V}n_{D}$ distribution.  This could be further
checked by comparing the inferred matter distributions with those
measured by weak lensing, rotation curves, and other emissions.

Finally, this proposal makes definite testable predictions for the
properties of the emitted spectra.  In particular, both the 511\keV\
spectrum and the diffuse\keV\ spectrum are produced on an
event-by-event basis, and are thus independent of the rate at which
the annihilation processes occur (the timescale for the emitting ``hot
spots'' to radiate is much shorter than the local collisional
frequency, so the nuggets do not ``integrate'' the effects of the
annihilations.)  The observed spectra should thus be largely
\emph{independent of the direction of observation.}\footnote{There may
  by a very small sensitivity to the local velocity of the incoming
  matter, but this is a small correction.}  Only the intensity should
vary as a function of the collision rate, and this should be
correlated with the visible/dark matter distribution as discussed
above.

\section{Conclusion:}
Our dark matter proposal not only explains many astrophysical and
cosmological puzzles, but makes definite predictions about the
correlations of the dark and visible matter distributions $n_{V}n_{D}$
with 511\keV, x-ray and gamma-ray emissions.  In addition, it
makes the definite prediction that the spectra of the emissions should
be virtually independent of the local environment.  Such correlations
and spectral properties would be very difficult to account for with
other dark matter candidates.  Future observations may thus confirm or
rule out this theory.  If confirmed, it would provide a key for many
cosmological and astrophysical secrets, and finally unlock nature of
dark matter.

As an aside, we mention that the mechanism for producing the
x-ray radiation discussed in the appendix is likely the
dominant mechanism for radiation from the surface of bare strange
quark stars, should they exist, though it seems to have been neglected
in the literature.

\acknowledgments The authors would like to thank Jeremy Heyl and Tom
Quinn for useful discussions, John Beacom, Francesc Ferrer, and Andrew
Strong for comments on the manuscript, the \textsc{pitp} program where
this project was started, the US Department of Energy for support
under Grant DE-FG02-97ER41014, and the Natural Sciences and
Engineering Research Council of Canada.

\ifJCAP{
  \bibliographystyle{hunsrt}
}{ 
  \bibliographystyle{h-physrev3}
}

\ifJCAP{
\appendix
}{ 
\appendix
}
\section{Emission of keV Radiation}
The goal of this appendix is to explain the process through which
proton annihilation inside the dark antimatter nuggets could release
the\keV\ scale photons observed \Chandra\ emissions.
The emission and spectrum of the 511\keV\ emission from
electron-positron annihilation is discussed in detail
in~\cite{Oaknin:2004mn,Zhitnitsky:2006tu,Lawson:2007kp}.

Although the details are somewhat complicated and model dependent, we
shall argue that the characteristics of this emission are most
strongly governed by the properties of the positron plasma with the
emission scale being set by the plasma frequency near the surface of the
nuggets.  As such, the emission is largely insensitive to the model
dependent details of the nugget core.

The basic idea is that protons could penetrate into the antimatter
nuggets and annihilate near the surface.  The 2\GeV\ of energy
released by this annihilation is ultimately transferred to high-energy
positrons that propagate out of the surface of the nuggets.  These
positrons are subsequently accelerated by the large electromagnetic
fields present in the electrosphere, and release their energy as
Bremsstrahlung radiation.  The scale and spectral properties of this
emission are determined by the properties of the electrosphere of the
nuggets where the\keV\ scale and approximately ``thermal'' nature of
the spectrum arises quite naturally.

\subsection{Structure of Nugget Surface}
Here we briefly discuss the properties of the antimatter nuggets
surface to set the scales.  The radius of the nuggets depends on the
mass, but must be larger than $R > 10^{-7}$~cm at the lower limit
$|B|>10^{20}$ set by terrestrial non-detection, but are most likely
several orders of magnitude larger.  The quark-matter core of the
nuggets ends sharply on a fm scale as set by the nuclear scale.  Near
the surface, as the density falls, the quark matter will be charged
due to the relatively large mass of the strange quark $m_{s}\sim
100\MeV$~\cite{Alcock:1986hz,Kettner:1994zs,Madsen:2001fu,Usov:2004iz}.
(The matter may be charged throughout depending on the exact nature of
the phase, but this does not affect our analysis.)  Charge neutrality
will be maintained through beta-equilibrium, which will establish a
positron chemical potential $\mu_{e^{+}} \simeq 10\MeV$.  (The precise
value depends on specific details of quark matter phase and may range
from a few\MeV\ to hundreds of\MeV, and is about an order of magnitude
less than the quark chemical potential $\mu_{q}\simeq
10\MeV$~\cite{Alcock:1986hz,Steiner:2002gx}.)  This will induce a thin
but macroscopic ``electrosphere'' of positrons surrounding the quark
matter core in the transition region as $\mu_{e^{+}}\rightarrow 0$ in
the vacuum.

The structure of this electrosphere has been considered for quark
matter~\cite{Alcock:1986hz,Kettner:1994zs} and the existence of this
``transition region'' is a very generic feature of these systems.  It is
the direct consequence of the Maxwell's equations and chemical
equilibrium.  The region is called the electrosphere, emphasizing the
fact that quarks and other strongly interacting particles are not
present.  In the case of antimatter nuggets the ``electrosphere''
comprises positrons.

The variation of this chemical potential $\mu_{e^{+}}(z)$, and the
associated electric field $E(z)$ as a function of distance from the
surface of the nugget $z$, may be computed using a mean-field
treatment of the Maxwell
equations~\cite{Alcock:1986hz,Kettner:1994zs,Usov:2004kj}.  For
example, in the relativistic regime, one has~\cite{Usov:2004iz}
\ifJCAP{
  \begin{eqnarray}
    \label{eq:field}
    \mu_{e^{+}}(z) &=\sqrt{\frac{3\pi}{2\alpha}}\frac{1}{(z+z_{0})},\qquad
    z_{0} &= \sqrt{\frac{3\pi}{2\alpha}}\frac{1}{\mu_0},
    \nonumber\\
    E(z) &= \frac{1}{e}\frac{\mathrm{d}\mu_{e^{+}}(z)}{\mathrm{d}z},
  \end{eqnarray}
}{ 
  \begin{align}
    \label{eq:field}
    \mu_{e^{+}}(z) &=\sqrt{\frac{3\pi}{2\alpha}}\frac{1}{(z+z_{0})},&
    z_{0} &= \sqrt{\frac{3\pi}{2\alpha}}\frac{1}{\mu_0},
    \nonumber\\
    E(z) &= \frac{1}{e}\frac{\mathrm{d}\mu_{e^{+}}(z)}{\mathrm{d}z},
  \end{align}
} where $\mu_0\equiv \mu_{e^{+}}(z=0) \sim 10\MeV$ is the chemical
potential realized in the nugget's bulk.  The corresponding results
can be obtained outside of the relativistic regime, but they do not
have a simple closed form.

Near the surface of the nugget, the typical scale is $z_{0} \sim
10^{-11}$~cm $\approx 100$~fm, thus, at a typical distance of
$1000 z_{0} \sim 10^{-8}$~cm $\approx 10^{5}$~fm, the chemical
potential will be of the\keV\ scale.  The electrosphere will extend well
beyond this, but this is the most important scale for the present
discussion.
\subsection{Excitations and Interactions}
The physics of radiation is dominated by the low-energy degrees of
freedom present in the quark matter and in the electrosphere.  Within
the nuggets, the picture is of some charged phase of
colour-superconducting matter with a gap of some $100\MeV$, with a
free Fermi-gas of positrons to maintain neutrality.  (See
\cite{Alford:2006fw} for a review of colour superconducting
properties.)
\begin{description}
\item[Quarks and Gluons:] The fundamental excitations in quark matter
  at sufficiently high densities are the quarks an gluons.  These
  excitations are coloured and strongly interacting, but have ``masses''
  on the order of the superconducting gap $\sim 100\MeV$.
  Thus, although the initial annihilation will produce these, the
  energy will ultimately be transferred to lower energy degrees of freedom.
\item[Pseudo--Nambu-Goldstone Bosons (Mesons):] The spontaneous
  breaking of chiral symmetry by colour-superconductors gives rise to
  low-energy pseudo--Nambu-Goldstone modes with similar quantum
  numbers of mesons (pions, kaons etc.).  These objects, however, are
  collective excitations of the colour-superconducting state rather
  than vacuum excitations.  The finite quark masses explicitly break
  the chiral symmetry, giving rise to these
  ``pseudo''--Nambu-Goldstone modes on the order of $10\MeV$
  (see~\cite{Son:1999cm,Son:2000tu} for example).  These modes are all
  strongly interacting and include both electrically charged
  ($K^{\pm}$, $K$, $\bar{K}$, $\pi^{\pm}$) and neutral ($\pi^{0}$,
  $\eta$, $\eta'$) modes.
\item[Nambu-Goldstone Phonon:] There is one truly massless
  Nambu-Goldstone mode associated with spontaneous baryon number
  violation.  This mode, however, carries neither colour nor electric
  charge.  Thus, it is extremely weakly coupled and plays an
  insignificant role in the radiative processes we consider here.
\item[Positrons:] The positrons exist as a Fermi liquid with chemical
  potential $\mu\sim 10\MeV$ and Fermi energy $p_{F}$.
  ($p_{F}^2=\mu^2-m_{e^{+}}^2 \approx \mu^2$.)  The
  total density of positrons is $n_{e^{+}}^{\rm tot} =
  p_{F}^3/(3\pi^2)$, however, most of the states are Pauli blocked
  deep within the Fermi surface.  For a given energy range $|E| <
  \omega$, we can estimate the number of available quasiparticle
  modes:\footnote{In principle, we should include many-body
    corrections in these calculations.  For example, the effective
    mass of the quasiparticles is $m_{\rm eff} = m_{e}/2 +
    \sqrt{m_{e}^2/4 + \alpha\mu^2/(2\pi)}$ \cite{1992ApJ...392...70B},
    but for $\mu = 10\MeV$, this is only a $30\%$ correction.  We
    neglect these corrections here for our order of magnitude
    estimates.}
  \begin{equation}
    \label{eq:n_e}
    n_{e^{+}} \simeq \frac{2\omega \mu p_{F}}{\pi^2}
    \approx \frac{2\omega \mu^2}{\pi^2}.
  \end{equation}
\item[Photon:] The electromagnetic interaction is mediated by a
  photon, but in the Fermi liquid, the interaction is screened and the
  photon has a plasma-frequency $\omega_{p}$ (see for
  example~\cite{Alcock:1986hz}):
  \begin{equation}
    \label{eq:w_p}
    \omega_p^2\simeq \frac{4\alpha}{3\pi}\mu^2.
  \end{equation}
  For $\mu_{q} \sim 100\MeV$, $\omega_{p}\approx 5\MeV$.  This has several
  implications: 1) low energy photons do not propagate freely, 2) the
  $1/q^2$ singularity of the Coulomb interaction is softened $1/q^2
  \rightarrow 1/(q^2 + \omega_{p}^2)$.  Electromagnetic interactions
  are thus dominated by momenta transfer of $q\sim \omega_{p}$.
  We also consider the mean free path for high energy photons which
  can be estimated by considering the Compton scatting cross-section
  $\sigma_{e^{+}\gamma}$ in the positron's rest frame (approximated
  for $\omega \gg m_{\rm eff}$):
  \begin{equation}
    \label{eq:l_gamma}
    \sigma_{e^{+}\gamma} \simeq \frac{\pi\alpha^2}{\omega m_{e}}
    \ln\left(\frac{\omega}{m_{e}}\right), \qquad\qquad
    l_{\gamma}^{-1} \simeq \sigma_{e^{+}\gamma}n_{e^{+}}(\omega_{p}).
  \end{equation}
\end{description}
From equations (\ref{eq:n_e}) and (\ref{eq:w_p}) we can estimate the
mean-free path for charged particles in the positron gas.  The
cross-section between the charged meson modes (for example, the
quark-matter equivalent of the $\pi^{+}$) and positrons is
\begin{equation}
  \label{eq:l_NG}
  \sigma_{e^{+}\pi^{\pm}} \simeq 2\pi\frac{\alpha^2}{\omega_{p}^2},
  \qquad\qquad
  l_{\pi^{\pm}}^{-1} \simeq
  \sigma_{e^{+}\pi^{\pm}}n_{e^{+}}(\omega_{p}).
\end{equation}
The interaction here is dominated by momentum transfer of $\omega_{p}$
due to the screened photon.  This has typical values of about
$l_{\pi^{\pm}} \sim 10^{-9}$~cm for $\mu_{e^{+}} \sim 10\MeV$.
\subsection{Energy Transfer}\label{sec:energy-transfer}
Now we consider how energy is transferred from proton annihilation
events to radiation.  Protons impinging upon dark antimatter nuggets
will penetrate some characteristic depth $d\ll R$ much less than the
radius of the nuggets.  The depth will be set by the nuclear
interaction scale: For example, if the annihilation
$\text{p}^{+}\text{p}^{-}$ cross-section is taken to be the vacuum
value, the proton will have a lifetime of only 2~fm/$c$.  It has been
argued~\cite{Mishustin:2004xa} that in nuclear matter, the lifetime
could be considerably larger (possibly even 10--30~fm/$c$) due to the
coherence required for the annihilation.  The lifetime may be even
longer in colour superconducting antimatter where the quantum numbers
for all three anti-quarks must be correlated for a successful
annihilation with incoming proton, perhaps even as large as
100~fm$/c$.

In any case, on the scale of the nuggets, the annihilations take place
on the order of $h\sim 100$~fm or so from the surface, which is still within
the regime where the quark matter is charged.  Thus, the annihilations
take place in a region of quark matter where there positrons are
present to maintain beta-equilibrium and neutrality.  These positrons
are the lightest modes that couple strongly and will thus ultimately
carry most of the annihilation energy as we shall now explain.

The proton annihilation will initially produce a handful of high
energy gluons.  These will stream away from the annihilation site.
About half of these will stream towards the surface of the nuggets,
and will ultimately transfer their energy to high-energy positrons.
The other half, however, will proceed deep within the nuggets, and
will most likely thermalize.  Thus, only about half of the initial
2\GeV\ per annihilation will be converted into energetic positrons
that propagate to the surface of the nuggets.  The remaining energy
will thermalize and subsequently be thermally radiated by the nuggets
at very low temperatures.  We account for this and other losses
through a factor $g \lesssim 1/2$ which ultimately appears in
our energy budget~(\ref{eq:f}).  We shall not further discuss this low
temperature thermal radiation in this work, though it may have further
observational consequences that need to be explored.

The remaining particles stream toward the surface of the nuggets and
have a typical distance $d\sim \sqrt{2Rh} \sim 10^{-8}$~cm to travel
before they hit the surface.  At the annihilation point, with
probability $\alpha/\alpha_{s}$, one of the decay products will be a
high energy photon with $\omega \sim 1\GeV \gg \omega_{p}$.  Such
photons have a long mean-free paths, on the order of
$l_{\gamma}^{1\GeV}\sim 10^{-8}$~cm, and can directly leave the
nuggets.  These are the photons considered in~(\ref{eq:aEGRET_comp_N})
and (\ref{eq:aEGRET_comp_E}) and should be observable by \EGRET.

Only the photons directly produced by direct annihilation will have
enough energy to escape.  Photons from additional processes, such as
the decay of the neutral pseudo--Nambu-Goldstone bosons ($\pi^0,\eta
\rightarrow 2\gamma$), will have energies $\sim 50\MeV$.  Not only are
these processes unlikely to occur near the surface, (the typical width
is $\Gamma\sim \alpha^2 m_{NG}^3/(64\pi^3 f_{\pi}^2)$ corresponding to
a huge mean free path\footnote{Although the pseudo--Nambu-Goldstone
  modes have similar quantum numbers to the vacuum mesons, their
  masses differ by an order of magnitude from their vacuum
  counterparts.  Coupled with the sharp quark-matter/vacuum interface,
  they thus cannot easily leave the quark matter, and will simply
  reflect off of the surface, ultimately re-scattering their energy
  into other particles.}) but the mean free path for these photons is
much shorter $l_{\gamma}^{50\MeV} \sim 10^{-10}$~cm, so they will be
absorbed long before they can escape.  As a result, we do not include
this energy in our estimates (\ref{eq:aEGRET_comp_N})
and~(\ref{eq:aEGRET_comp_E}).

The remaining gluons will rapidly decay into the
pseudo--Nambu-Goldstone collective modes, distributing the energy down
to their mass scale of $10\MeV$ or so.  As estimated
in~(\ref{eq:l_NG}), the charged modes scatter quite readily with
positrons, transferring energies on the order of $\omega_{p} \sim
5\MeV$.  These energetic positrons will travel a similar distance
before scattering, (further distributing the energy down to this scale
if it is above $\omega_{p}$).  Unlike the collective modes, however,
the positrons are not bound to the nuggets.

Thus, through this mechanism, some fraction $g \lesssim 1/2$ of the
initial $2\GeV$ annihilation energy will ultimately be transferred to
positrons of energy $2 \sim 5\MeV$ which stream out of the nuggets
into the electrosphere.  A small fraction $\alpha/\alpha_{s}$ of the
events may also produce $\GeV$ scale photons that directly escape.
The rest of the energy is transferred deep into the nuggets,
ultimately thermalizing and emitted at very low temperatures (to be
discussed elsewhere~\cite{Forbes/Zhitnitsky:2007b}).



\subsection{Bremsstrahlung Radiation}
The remaining emission processes of the dark antimatter nuggets
results from the physics of these 100 or so energetic positrons
($E\sim 5$\MeV) that enter the electrosphere of the nuggets.  (For
simplicity, we assume that all of the positrons have an energy of
$5\MeV$, being set by the plasma frequency of the photons that excited
them.  A detailed calculation of the spectrum would need to account
for the scatter in this value.)  As these energetic positrons move
through the strong electric field $E(z)$, they immediately start to
emit photons.  For a simple estimate of the timescale over which the
energy is released, we treat these positrons classically.  In this
case, the total instantaneous radiated power is given
by~\cite{Jackson:1999} \ifJCAP{
\begin{eqnarray}
 P_{\gamma} &= \frac{2\alpha\gamma^6}{3}\left[
   (\dot{\vec{v}})^2
   -(\vec{v}\times\dot{\vec{v}})^2\right],\qquad
 \gamma&\equiv\frac{1}{\sqrt{1-v^2}},
\end{eqnarray}
}{
\begin{align}
 P_{\gamma} &= \frac{2\alpha\gamma^6}{3}\left[
   (\dot{\vec{v}})^2
   -(\vec{v}\times\dot{\vec{v}})^2\right], 
 &\gamma&\equiv\frac{1}{\sqrt{1-v^2}},
\end{align}
}
which we express in terms of the local electric $\vec{E} $ and
magnetic $\vec{B}$ fields evaluated at the position of the
particle~\cite{Jackson:1999},
\begin{equation} 
  \label{eq:P}
  P_{\gamma} = \frac{2\alpha^2\gamma^2}{3m^2}\left[
    (\vec{E}+\vec{v}\times \vec{B} )^2 -(\vec{v}\cdot \vec{E})^2\right].
\end{equation}
Now we can easily estimate the energy radiated per unit time by a
positron in the background electric field (\ref{eq:field}) as
\begin{equation} 
 \label{P1}
 P_{\gamma}\sim \mu^2\cdot\frac{4\alpha^2\gamma^2}{9\pi}\cdot
 \left( \frac{\mu^2 }{m^2}\right)\cdot \left(\frac{z_{0}}{z+z_{0}}\right)^4.
\end{equation}
The total time for the positron with energy $\epsilon = 5$\MeV\ to
release its energy in the form of synchrotron radiation is
\begin{equation} 
  \label{tau}
  \tau\sim \frac{ \epsilon}{P_{\gamma}} \sim 
  \frac{\epsilon}{\mu^2}\left( \frac{ m^2}{\mu^2}\right)
  \frac{9\pi}{4\alpha^2\gamma^2}  \left(\frac{z+z_{0}}{z_{0}}\right)^4.
\end{equation}
Numerically, if we take $\mu\sim 10\MeV$ and assume that most
efficient radiation is happening at a typical $z$ of a few times
$z_{0}$ with an average $\gamma \sim \epsilon/(2m)$, we arrive at the
estimate $c\tau \sim 10^{-8}\text{cm}$.

In order for a photon to be efficiently emitted, its energy must be
greater than the local plasma frequency $\omega_{p}$.  This is on the
order of $5\MeV$ at the surface of the nugget, but rapidly falls off
with $\mu_{e^{+}}(z)$ so that by the time the photons have been formed
at a distance on the order of $10^{-8}$~cm, the plasma frequency has
dropped to the $\keV$ scale.  This is consistent with fact that the
wavelength of the emitted photons cannot be larger than length over
which the photons are emitted $z=c\tau \sim 10^{-8}$~cm, corresponding
to a minimum frequency $\omega \geq 2$\keV.  This is not a lower bound
on the spectrum, however, because additional scattering may also
occur, distributing some of the initial energy of the positrons and
extending the range of emission further into the electrosphere.  This
will soften the emission, but the characteristic energy scale will be
$\keV$.

\subsection{Can  Bremsstrahlung Mimic the Thermal Plasma
  Emission  Spectra?}
\label{sec:thermal-bremsstralung}
In this section, we show that the resulting $\keV$ Bremsstrahlung
spectrum might be identified with the $8\keV$ thermal plasma emission
spectrum used to fit the \Chandra\ data.  The most prominent feature
of a thermal plasma emission is the presence of the exponential factor
which suppresses emission with $\omega\gg T$,
\begin{equation} 
  \label{eq:omega}
  \frac{{\rm d}P_{\gamma}(\omega)}{{\rm d}\omega} \sim e^{-\omega/T}.
\end{equation}
The characteristic spectrum is thus very broad, (almost flat up to a
mild $\ln\omega$ dependence which has been omitted from
(\ref{eq:omega})) with a sharp cutoff at $\omega_c\sim T$.  This
exponential cut off is related to the minimum energy
$mv_{\text{min}}^2/2 \sim T$ required to emit a photon with frequency
$\omega$ by thermal plasma.  We note that the cutoff has not been
directly observed because the detector band is limited at $8\keV$,
thus the spectrum might extend to higher energies (10~\keV or so).
(See~\cite{Muno:2004bs} for details).

Although we cannot present a detail model-inde\-pendent calculation of
the spectrum, we shall argue that the qualitative features of the
observed thermal spectrum are reproduced by field induced
bremsstrahlung emission.  In particular, there is a critical frequency
$\omega_{c}$ above which emission is strongly suppressed, and the
spectrum is very flat for low energies $\omega \ll \omega_{c}$.  This
behaviour is in sharp contrast with a black body spectrum, for
example.

A quantitative treatment of the problem is difficult because the
photon's emission must be calculated with a fully quantum mechanical
treatment that takes into account the many-body effects and positron
degeneracy.\footnote{Photon emission in a dilute system
  will be suppressed by $\alpha^3$ while equation (\ref{eq:P})
  explicitly demonstrates $\alpha^2$ behavior.  The enhancement is due
  to the coherent electric field (\ref{eq:field}) which has $E\sim
  \frac{1}{e}$ behavior.  This means that, in dense systems, the
  bremsstrahlung emission cannot be treated as a series of independent
  events ($\alpha^3$ behavior), but must be treated as a coherent
  emission.  The relevant parameter here is the ratio between the
  distance between electrons ($\mu^{-1}$) and the typical energy
  transfer in each event ($\omega$).  The number of particles
  participating in the emission is very large, ($(\mu/\omega)^3\gg
  1$), therefore, the standard Bremsstrahlung emission technique
  cannot be used for our system.  A precise calculation of the
  spectrum is beyond the scope of the present work.}  Ultimately this
will be required to confirm or rule out any particular model.  A
qualitative picture for the spectrum, however, follows from simple
arguments.  Consider the motion of a relativistic classical particle
in the background electric field (\ref{eq:field}).  The quantum
mechanical emission of ``soft'' photons---photons whose energies are
small compared with the typical interaction energy scale---is almost
identical to that obtained by classical analysis (see for example
\cite{Jackson:1999}).  Furthermore, the most important contribution
comes from the small part of the classical trajectory with the largest
instantaneous curvature.  It is exactly this region that determines
the critical value for $\omega_{c}$ and the qualitative features of
the spectrum.

This key segment of the classical trajectory leads to the following
emission spectrum~\cite{Jackson:1999},
\begin{equation}
  \label{spectrum}
  \frac{{\rm d}P_{\gamma}(\omega)}{{\rm d}\omega}\sim
  \frac{\omega}{\omega_c}
  \int^{\infty}_{\omega/\omega_c}K_{5/3}(x){\rm d}{x},  
\end{equation}
with the asymptotic behaviors
\begin{equation}
  \left.\frac{{\rm d}P_{\gamma}(\omega)}{{\rm d}\omega}
  \right|_{\omega\ll \omega_c}\sim \omega^{1/3}, \qquad\qquad
  \left.\frac{{\rm d}P_{\gamma}(\omega)}{{\rm d}\omega}
    \right|_{\omega\gg\omega_c}\sim  e^{-\omega/\omega_c}.
\end{equation}
The main feature of the spectrum is that it is very broad and flat at
frequencies below $\omega_c$ with a sharp cut off for $\omega\geq
\omega_c$. In particular, when $\omega/\omega_c$ changes by two orders
of magnitude ($10^{-2} <\omega/\omega_c < 1$), the intensity changes
only by factor 2. These features are very similar to those of a
thermal plasma emission spectrum where the role of $T$ is played by
$\omega_c$.
    
To estimate for $\omega_c$, we note that the typical time scale for
the changes in the system is $\delta t\sim \epsilon / eE$ where
$\epsilon\sim \mu_0+\epsilon_0$ is a total initial energy of the
particle at $t=0$.  This parameter determines a curvature of the
classical trajectory.  The critical frequency of the emission
$\omega_c$ in such a motion is $\omega_{c}\sim (\delta t)^{-1}$ can be
estimated for the classical field (\ref{eq:field}) as follows,
\begin{equation}
  \label{omega_c}
  \omega_c\sim  \frac{eE}{\mu_0+\epsilon_0}\sim \sqrt{\frac{2\alpha}{3\pi}}\left(\frac{z_0}{z+z_0}\right)^2
  \frac{\mu_0^2}{\mu_0+\epsilon_0}. 
\end{equation}
Numerically this is on the order of $30\keV$ or so.  Without a full
quantum calculation, this is only an order of magnitude estimate, but
it is is consistent with our estimate for minimal frequency $\omega
\geq 2\keV$ obtained above.


We note here that this mechanism, whereby the moving positrons radiate
as they are accelerated by the strong electric fields in the
electrosphere, is likely the dominant mechanism radiating energy from
the surface of strange quark matter such as strange stars.  In this
case, thermally excited electrons (strange stars would be matter
rather than antimatter) would accelerate in the fields of the star's
electrosphere, producing a similar radiation.  This mechanism has
apparently been overlooked in the
literature~\cite{Prakash-Page-Usov-PC:2007}.  Our estimates here
provide a starting point, but a full quantum mechanical computation
should be done to properly account for rate at which thermal energy
may be radiated in these systems.
  
\section{Cosmic Rays Impinging on Nuggets}
It has been suggested that much of the \EGRET\ emission could result
from cosmic rays (see for example~\cite{Strong:2004de}).  This would
affect the constraints on the $a_{\EGRET}$ as we have discussed, but
raises another question: Could interactions between cosmic rays and
nuggets have an observable consequence in the core of the Galaxy?  The
typical energy released by a cosmic-ray proton annihilating on a
nugget would be $30\GeV$ rather than the $1\GeV$ released in the
low-energy annihilations we have considered so far.  In addition ,
there would be no suppression factor $f\sim 10^{-2}$ for these high
energy events, thus, if there were a significant number of collisions,
there would be an observable effect.  As we shall see, the rate of
collisions is very small, so that the net energy released is four
orders of magnitude less than our total budget, so we conclude that
such processes may be neglected.

We start from analysis of the conventional and optimized
models~\cite{Strong:2004de} for proton spectra as the most important
portion of cosmic rays.  Data presented in~\cite{Strong:2004de}
suggests the cosmic-ray proton flux $I(E)$ to be about
\begin{multline}
  \label{CR}
  E^2 \frac{{\rm d} I(E)}{{\rm d}\Omega {\rm d}E} \sim 2
  \times 10^{-1}\frac{\GeV}{\text{cm}^2 \cdot \text{ s} \cdot \text{
      sr}}\\
  \text{where }
  2\GeV \leq E \leq 30\GeV,
\end{multline}
which corresponds to a total integrated flux
\begin{multline}
  \label{CR1}
  I=\int  \frac{{\rm d} I(E)}{{\rm d}\Omega dE} {\rm d} \Omega {\rm d}E 
  \sim \\
  \sim 2\times10^{-1}
  \int \frac{{\rm d} E}{E^2}  \cdot \frac{\GeV\cdot 4\pi
  }{\text{cm}^2\cdot \text{ s} \cdot \text{ sr}}
  \sim 1
  \frac{\text{proton}}{\text{cm}^2\cdot \text{ s}}.
\end{multline}
The corresponding cosmic-ray proton flux (\ref{CR1}) can be used to
estimate the total number of the collision events $\Phi_{CR}$ of
cosmic-ray protons with the nuggets along the line of sight of the
observation,
\begin{equation}
 \label{CR2}
 \Phi_{CR}\sim \int {\rm d}r \int_{\Delta \Omega} {\rm d}\Omega \cdot
 n_{DM}(l)\cdot  I \cdot (4\pi R_0^2),
\end{equation}
where $ \Delta \Omega $ is the solid angle observed, and the integral
$\int {\rm d}r$ is performed over the line of sight of the observation.  We
assume that all cosmic-ray protons hitting the nuggets will
annihilate, and that the total integrated flux of cosmic-ray protons,
$I$---which enters equation (\ref{CR2}) and is estimated in (\ref{CR1})---
does not varies much with distance.  (This would over-estimate the
flux.)

The number density of the nuggets is approximately $n_{DM}(r) =
\rho_{DM}(r) / M_{DM}$ with $M_{DM}$ being a typical mass of the
nugget.  Numerically, $\Phi_{CR}$ is very small, about seven orders of
magnitude smaller than the typical rate of collision between visible
protons and nuggets.  Such a suppression can be easily understood:
When calculating the number of collision events between visible matter
protons and nuggets along the same line of sight, one should replace
$I$ in formula (\ref{CR2}) by $n_B\cdot v$.  Assuming $n_B\sim
1$~cm$^{-3}$, and taking a typical $v/c\sim 10^{-3}$, one finds this
seven orders of magnitude difference in the annihilation rate, which
more than compensates for the enhanced annihilation rates (no factor
of $f \sim 1/100$) and extra energy ($30\GeV$ rather than $1\GeV$)
provided by these events.
\end{document}

length :-> "cm"
energy :-> "MeV"

rho0=5GeV/(fm^3)
R[B,rho0=5GeV/(fm^3)]:=1GeV*B/(4/3*pi*rho0))^(1/3)

m_eff[mu_e]:=m_e/2 + sqrt[m_e^2/4 + alpha/2/pi*mu_e^2/c^4]

wp[mu]:=sqrt[4*alpha/3/pi*mu^2]

n_e[w,mu_e]:=2/pi^2*mu_e*sqrt[mu_e^2-(m_eff[mu_e]*c^2)^2]*w/(hbar*c)^3

s_eNG[mu_q,mu_e]:=2*pi*alpha^2/wp[mu_q]^2*(hbar*c)^2

s_eph[w,mu_e]:=pi*alpha^2/w/(m_eff[mu_e]*c^2)*ln[2*w/(m_eff[mu_e]*c^2)]*(hbar*c)^2

l_NG[mu_q,mu_e]:=1/(s_eNG[mu_q,mu_e]*n_e[wp[mu_q],mu_e])

l_ph[w,mu_q,mu_e]:=1./(s_eph[w,mu_e]*n_e[wp[mu_q],mu_e])

z0[mu]:= sqrt[3*pi/2/alpha]/mu*hbar*c

mu_z[f,mu_e]:=sqrt[3*pi/2/alpha]/((f+1)*z0[mu_e])*hbar*c

E[f,mu_e]:=mu_z[f,mu_e]/((f+1)*z0[mu_e])

tau[ee,mu_e,f]:= ee/mu_e^2*(m_eff[mu_e]^2*c^4/mu_z[f,mu_e]^2)*9*pi/4/alpha^2/(ee/2/m_eff[mu_z[f,mu_e]])^2*(f+1)^4*hbar*c^5

B=1e25
h=100fm
mu_q=100MeV
Ms=120MeV
mu_e=Ms^2/4/mu_q
f_pi=0.2*mu_q
m_NG=10MeV
println["B = " + B]
println["R = " + R[B]]
println["Ms = " + Ms]
println["mu_q = " + mu_q]
println["mu_e = " + mu_e]
println["z0 = " + z0[mu_e]]
println["wp_q = " + wp[mu_q]]
println["wp_e = " + wp[mu_e]]
println["l_NG = " + l_NG[mu_q,mu_e]]
println["h = " + h]
println["d = " + sqrt[2*R[B]*h]]
println["l_ph(GeV) = " + l_ph[1GeV,mu_q,mu_e]]
println["l_ph(100MeV) = " + l_ph[10MeV,mu_q,mu_e]]
println["Gamma = " + alpha^2 m_NG^3/(64*pi^3*f_pi^2)]
println["1/Gamma = " + 1/(alpha^2 m_NG^3/(64*pi^3*f_pi^2))*hbar*c]
println["tau(z=3.5*z0) = " + tau[5MeV,mu_e,3.5]]
println["w_c(z=3.5*z0) = " + (E[3.5,mu_e]/(mu_e+5MeV)*c*hbar->"keV")]
